\documentclass{acm_proc_article-sp}

\usepackage{graphicx}

\makeatletter
\def\@copyrightspace{\relax}
\makeatother

\begin{document}

\title{Monitoring Large-Scale Cloud Systems with Layered Gossip Protocols }

\numberofauthors{2} %  in this sample file, there are a *total*

\author{
\alignauthor Jonathan Stuart Ward \\
	\affaddr{School of Computer Science} \\
	\affaddr{University of St Andrews, Scotland}\\
	\email{jonthan.stuart.ward@st-andrews.ac.uk}
\alignauthor
Adam Barker\\
	\affaddr{School of Computer Science}\\
	\affaddr{University of St Andrews, Scotland}\\
	\email{adam.barker@st-andrews.ac.uk}
}

\maketitle
%\date{30 July 1999}
%A category including the fourth, optional field follows...
%\category{D.2.8}{Software Engineering}{Metrics}[complexity measures, performance measures]

%\terms{Cloud Computing, Aggregation}

\begin{abstract}

In this paper we propose the development of a cloud monitoring suite to provide scalable and robust lookup, data collection and analysis services for large-scale cloud systems. We propose a multi-tier architecture using a layered gossip protocol to aggregate monitoring information and facilitate lookup, information collection and the identification of redundant capacity. This enables monitoring to be done in-situ without the need for significant additional infrastructure to facilitate monitoring services. \end{abstract}

\section{Introduction}
Cloud computing has become a common means for the deployment of large-scale scientific 
and enterprise applications. Elasticity is one of the key properties central to the appeal of 
cloud computing, allowing for large-scale systems to be rapidly provisioned and decommissioned 
on-demand within a short period of time. With the immense benefits of this paradigm comes a number of challenges, amongst these is the challenge of monitoring. 

Monitoring large-scale distributed systems is challenging. It requires substantial
engineering effort to identify pertinent information and to obtain, store and
process that information in order for it to become useful. Despite the
difficulty, monitoring is a crucial component of any large-scale systems
deployment. Effective monitoring helps eliminate performance bottlenecks,
security flaws and is instrumental in helping engineers make informed decisions
about how to improve current systems and how to build new systems.
Existing monitoring systems are predominantly intended for monitoring physical
servers which are not usually prone to rapid change  which
occurs through elasticity~\cite{Ward}.  

Current monitoring
systems commonly rely on a central server, or set of servers to pull data from a
static pool of monitored servers. This design has two significant issues: (i)
centralised data collection substantially increases in cost and decreases in
speed as the size of the system increases and (ii) requires reconfiguration to 
add and remove individual servers. This approach is not ideal for monitoring large
scale cloud based systems.  Monitoring systems which are designed specifically
for the cloud are provisioned as a service; such services typically achieve
scalability by utilising conventional centralised tools supported by
significant backend infrastructure.

\newpage
In this paper we propose and evaluate an architecture for performing fully
decentralised data collection, analysis and monitoring of cloud virtual
machines (VMs). Our approach abandons the traditional monitoring concept in
lieu of a decentralised approach based upon a hierarchical gossip algorithm. By
leveraging this mechanism we propose an architecture which is highly scalable,
and attempts to eliminate the need for additional, dedicated monitoring infrastructure. 
It is through this approach that we intend to meet the requirements imposed 
by rapid-elasticity~\cite{Ward} and provide a monitoring system suitable for cloud computing.

\section{Architecture}
Gossip protocols operate analogous to the manner in which gossip spreads over
a social network. Gossip style communication is highly responsive, lacks any 
single point of failure, exerts minimal load on individual processes and is
extremely scalable when compared to other methods of group communication~\cite{Birman}. Gossip 
communication has been used to great effect as a basis for broadcast and multicast, database replication, failure detection and aggregation. Here we propose the use of a gossip
protocol to facilitate data collection and monitoring in large-scale cloud systems.

Our gossip protocol is layered in order to best exploit the topography of IaaS clouds and
reduce the volume of communication overhead. There are three layers: clouds, groups and 
VMs. The rationale for this hierarchy is rooted in the differences between intra and inter
cloud communication. Within clouds there is high bandwidth, low latency and the connection
is unmetered. This environment lends itself to the use of rapid, UDP based information dissemination.
Between cloud regions this is not as feasible, costs arising from latency and bandwidth metering force communication
to be performed in a slower, more reliable fashion. This therefore requires a slower, reliable 
protocol to synchronise state between regions.

Groups are created based upon the software installed in each VM. Group allocation uses a feature vector
to describe the primary applications which the VM operates. Groups are created based upon the 
proximity of VMs to one another within the feature space. This ensures that related VMs, e.g.,
all web servers or all Hadoop nodes are grouped together. This scheme is based upon the presupposition
that monitoring data from a VM is most relevant to other VMs which are similar to the first.

Each VM maintains a set of peers belonging to the same group and a list of contacts for
VMs in other groups in the same cloud region and different cloud regions. At every $T_{gossip}$ a VM selects
a number of other VMs from the same group propositional the size of the group, preferentially
selecting those with a lower communication latency. The VM will then gossip a message containing
a representation of its resource usage including: CPU, memory, disk and network utilisation
to each target. Each VM receiving the message selects additional targets and repeats this process.
This mechanism attempts to ensure that the resource usage of each VM is aggregated to its group.

\begin{figure}[h]
	\includegraphics[width=0.5\textwidth]{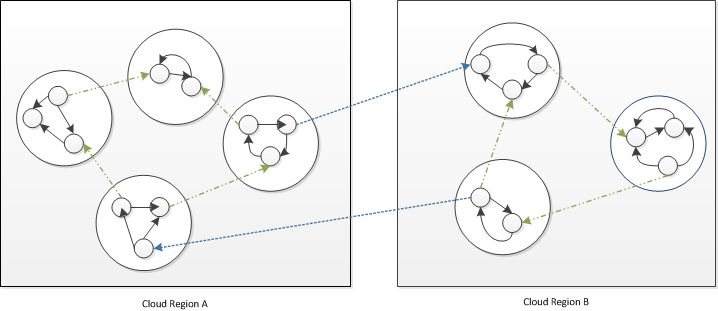}
	\caption{Monitoring architecture. Black lines denote frequent \emph{intra-group} communication, green denotes fast, less frequent \emph{inter-group} communication and blue lines denotes slow, infrequent \emph{inter-cloud} communication. Large circles denotes groups, small circles VMs.}
\end{figure}

Periodically, at a rate proportional to intra-group gossip, the aggregated resource usage 
of a group is gossiped to every other group by an agreed upon VM. Proportional to this communication
is inter-cloud communication whereby the aggregated resource usage of an entire
region is gossiped to every other region. Figure 1 illustrates this
communication hierarchy.  This ensures that every VM is aware of system wide
resource usage and serves as a basis for more complex monitoring. This design
has significant merit over existing designs which rely upon significant
dedicated infrastructure provided either by the user or by the cloud provider. When monitoring a heavily
loaded system, self monitoring is not feasible and requires some additional infrastructure to fulfil
all monitoring functions. Prior to this point, monitoring can be achieved
without additional infrastructure by distributing
monitoring functions throughout the monitored system.

\section{Use Case}
Our tool is intended to provide monitoring services for large-scale systems hosted on
an IaaS cloud such as Amazon EC2 or OpenStack without (or with a reduced) need for additional
monitoring infrastructure. A typical use case based on EC2 would involve a web application and 
associated infrastructure deployed over hundreds of virtual machines amongst three AWS regions. 
Rather than paying for a monitoring service or deploying additional servers to perform monitoring,
our architecture may be deployed over each VM to facilitate monitoring. The required software is
installed either at runtime or built directly into the VM image. Once VMs are started, a gossip based overlay network
will be constructed over the VM deployment and data collection and monitoring services will 
be provided through this overlay.

\section{Evaluation}
In order to evaluate our architecture we compare our architecture to flat gossip scheme and to
a centralised data collection architecture analogous to that used by a number of open source
monitoring systems, e.g., Nagios and Ganglia. We evaluate our architecture against other common architectures by way of 
software based simulation. Figure 2 shows the message rates of each communication scheme. The centralised
scheme employed by current monitoring systems is by far the most conservative due to its reliance on
unicast. This ensures that messages are kept to a minimum at the cost of the required computation to 
perform all monitoring functions in a single location. Of the two gossip strategies, the layered gossip
method utilised in our architecture has significantly lower message rates due to the grouping method. While
having still significantly higher messages rates than that of centralised monitoring systems, our architecture
attempts to exploit a tolerable middle ground where increased message rates are an acceptable trade-off in order
to achieve decentralised monitoring. While up to 74\% more messages are sent using our architecture than
conventional centralised tools, in a cloud setting such communication is entirely feasible given the presence
of high bandwidth, low latency connections. 
\vspace{-1\baselineskip}
\begin{figure}[h]
\begin{center}
   \includegraphics[width=0.45\textwidth]{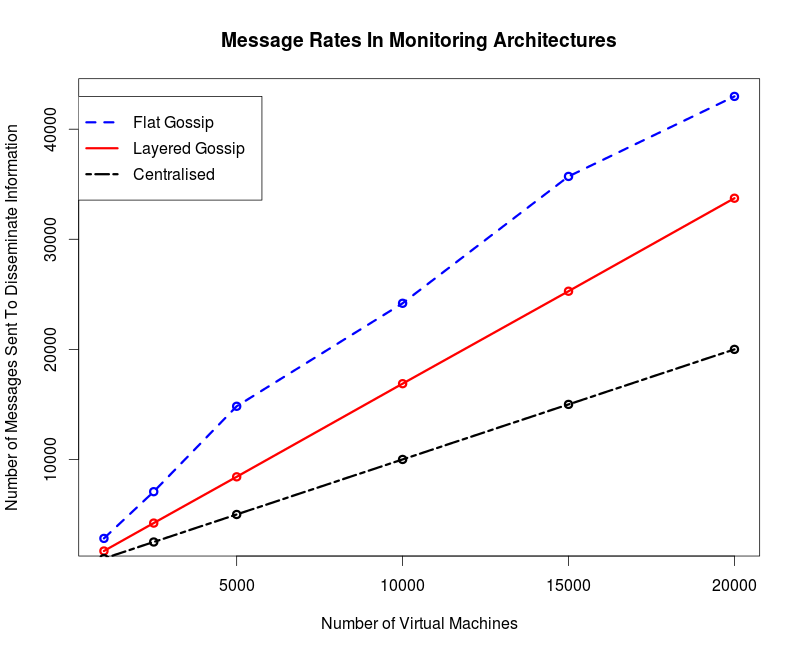}
   \caption{Communication overhead. }
   \end{center}
   \vspace{-2\baselineskip}
\end{figure}

\section{Conclusion and Future Research}
We have presented a framework for performing self-monitoring in a fully distributed fashion. This architecture
has significant merits over existing systems which extensively rely on additional infrastructure to performing
monitoring. Future research will investigate the functions which be be built upon the gossip infrastructure and how
a full, feature complete monitoring service can be provisioned.

\bibliographystyle{plain}
\bibliography{ref}

\begin{thebibliography}{1}

\bibitem{Birman}
Ken Birman.
\newblock {The Promise, and Limitations, of Gossip Protocols}.
\newblock {\em SIGOPS Oper. Syst. Rev.}, 41(5):8--13, October 2007.

\bibitem{Ward}
Jonathan~Stuart Ward and Adam Barker.
\newblock {Semantic Based Data Collection for Large Scale Cloud Systems}.
\newblock In {\em Proceedings of the fifth international workshop on
  Data-Intensive Distributed Computing}, DIDC '12, pages 13--22, New York, NY,
  USA, 2012. ACM.

\end{thebibliography}
\end{document}